# Epsilon Aurigae in Total Eclipse, 2010 – Mid-eclipse report

http://arxiv.org/pdf/1005.3738

*Robert E. Stencel*
*Meyer-Womble Observatory*
*© University of Denver Observatories, Denver CO 80208 USA*
*rstencel@du.edu*

**Abstract**

Epsilon Aurigae is a complicated binary star that undergoes optical eclipses every 27 years, including the present year. An update is given here on the array of photometric and spectroscopic observations underway, thanks to the eclipse observing campaign and its dedicated participants. In addition, breakthrough results have emerged from (1) infrared and ultraviolet spectral energy distribution observations, and (2) especially with interferometric imaging that revealed the long suspected dark disk in transit, plus (3) new optical spectra that are revealing substructure inside the disk itself. Implications of many of these observations is discussed, but as the eclipse data are still being collected, I anticipate that additional discoveries are still to come, throughout 2010, and beyond.

## Introduction

Epsilon Aurigae (ep-si-lon Awe-rye-gee) is an Algol-like, very long period eclipsing binary. The 27 year eclipse interval includes a nearly 2 year eclipse duration, wherein visual magnitude drops from 3.0 to 3.8. We've seen the start of the first eclipse of the millennium, during Aug. 2009, with totality reached during January 2010, predicted to last 15 months, through March 2011. The end of eclipse is expected during May/June 2011. For up to date details, see the on-line newsletter series (Hopkins, 2010). This eclipse will be only the seventh documented eclipse in history (2010, 1983, 1956, 1930, 1902, 1874 and 1847). For additional detail about epsilon Aurigae, see Stencel & Hopkins (2009), along with the book about the star by Hopkins & Stencel (2009), and feature articles in the May 2009 issue of Sky & Telescope, Oct. 2009 in Astronomy, and the Feb. 2010 issue of Astronomy Now, and elsewhere.

The bright eclipsing binary star, epsilon Aurigae, has long fascinated astronomers who could determine the nature of one companion star in the system, but not the other (Guinan and Dewarf, 2002). The crux of the problem was that the apparent early F supergiant star in this single lined spectroscopic binary, should have a comparably massive secondary. However, the secondary is vastly under-luminous for the presumed mass. Various models have been proposed, but the preferred model involves a disk of material obscuring a fairly normal companion star (Huang, 1965, 1974).

I am happy to report that an unprecedented series of observations are underway in photometric, spectroscopic, polarimetric and interferometric modes. Theory is advancing as well, as discussed below. Photometry and spectroscopy campaign efforts, light curves and eclipse timing results to date are discussed in a companion paper by Jeff Hopkins. New orbital solutions have appeared as well – see Stefanik et al. (2010) and Chadima et al. (2010).

## Development #1: The complete spectral energy distribution (SED).

As reported by Hoard, Howell and Stencel (2010, hereafter "HHS"), data have become available that span a wide spectral range, from the far-ultraviolet, through the visible range and out into the far-infrared. Because of calibration efforts, it has proven possible to combine these well calibrated data into a complete and self-consistent picture of the sources of light in epsilon Aurigae. Key to understanding this result is that interlocking requirements of distance and other constraints on F star diameter, drive us to these self-consistent conclusions.

Clearly the F star dominates much of the visible portion of the spectrum (Figure 1), and for the HIPPARCOS distance of 625pc, given the interferometrically determined diameter, 2.27 +/- 0.11 milli-arcsec, then the F star radius is 135 +/- 5 solar radii. The HIPPARCOS distance is not well determined, however. The energy distribution is well described by a 7750K surface temperature and log gravity = 1.0 atmospheric model. Subtracting this well-defined F star leaves two residual signals: In the far-ultraviolet, extra flux beyond the well-fit optical region, can be fitted by a B5V star, with a nominal mass equal to 5.9 solar masses, and radius of only 4 solar radii.

In the infrared, clearly an excess signal is present above the F star, and this is the cold disk previously measured by Dana Backman to have a 550K temperature. The total flux from the disk defines a luminosity and hence a total surface area. Using eclipse length as a width constraint (radius 3.8AU), the implied thickness is 0.9AU, consistent with the depth of eclipse – that is, the disk does not fully cover the F star.

Figure 1: The ultraviolet-optical-infrared spectral energy distribution of epsilon Aurigae.

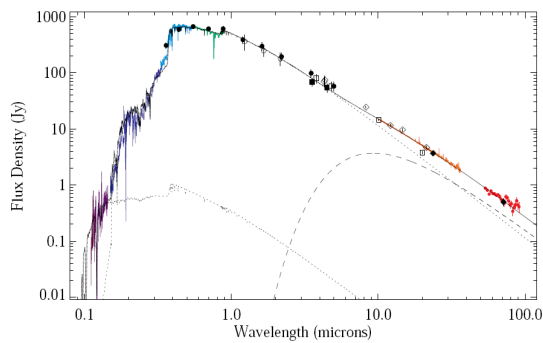

## Development #2 - Interferometric Imaging

A list of the major technology advances since the 1983 eclipse has to include computers and computer-aided positioning control with lasers (metrology). Both play a key role in interferometry – the ability to combine the light from separate telescopes and thereby achieve the diffraction-limited resolution of their effective apertures. For example, a single one-meter aperture telescope tends to be seeing-limited to about one-arcsec, although its theoretical diffraction limit is 1.22 (wavelength/aperture) = $7 \times 10^{-7}$ radians or 0.1 arcsec at V band. Two one-meter telescopes separated by 100 meters, interferometrically linked, can achieve 100 times that resolution, or 0.001 arcsec, or 1 milli-arcsec. The Sun at the distance of alpha Centauri would subtend 7 milli-arcsec. Interference fringes indicate the angular size of a resolved object on a line parallel to the separation of the two telescopes. With this two telescope technique, we have measured the pre-eclipse diameter of the F star in epsilon Aurigae to be 2.27 +/- 0.11 milli-arcsec (Stencel et al. 2008), which at the estimated 625 pc distance, means the star spans 135 solar diameters. A super-giant among stars!

With multiple telescopes interferometrically combined, it is also possible to achieve *phase closure* - which means being able to create images in the computer, subject to a number of assumptions. This technique has been used at the Mt. Wilson CHARA array and its MIRC imager to see the start of eclipse last autumn, as shown in Figure 2.

Figure 2: Ingress images of epsilon Aurigae, on the sub-milli arcsecond spatial scale, during ingress Nov and Dec 2009, from Kloppenborg et al. (2010).

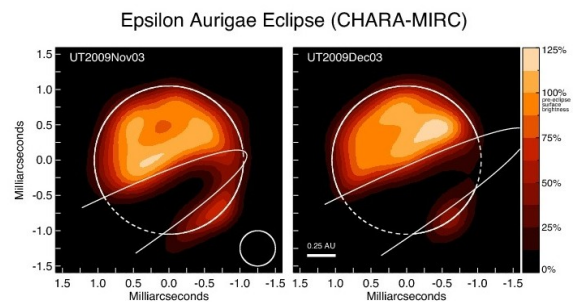

This pair of images, and hopefully more to follow, already demonstrate that the basic disk model (Huang, 1965) is correct, with modifications to account for disk shape. As reported by Kloppenborg, Stencel, et al. (2010), the measured direct motion of the disk, when combined with the well-known F star orbital motion, allows us to determine the mass ratio. Surprisingly, the F star mass is only 2/3 of the disk plus its contents. Adopting results of HHS (2010), the 5.9 solar mass central star in the disk implies an F star of 4 solar masses, or less – hardly a massive supergiant, but more likely in an unusual post-AGB state of evolution. Given the opacity and volume constraints on the disk, we estimate the disk mass to be of order Earth mass – similar to other *debris disks* – like those known to surround beta Pictoris and Vega.

Some of the questions resulting from direct detection of this disk are being considered as part of a doctoral dissertation by my thesis student, Brian Kloppenborg at the University of Denver. These include determining the state of evolution of the binary and its components; finding the cause of the out-of-eclipse light variations (hot spots on the F star? circum-binary clouds?); making sense of the changing lengths of eclipse ingress, totality and egress relative to the size of the F star and the disk; exploring the implications of disk structure and sub-structure (next section), and other questions. Does a disk that forms after a mass transfer event always surround the more massive star in the resulting system? Does the resulting lightweight star, if convective, end up with tidally-raised spots?

## Development #3 - Spectroscopic "imaging" and disk sub-structure.

As mentioned, among the campaign observers are those pursuing spectroscopic monitoring. One important result has already emerged thanks to Robin Leadbeater and his Three Hills Observatory in Cumbria, north of England. His steady recording of spectra, including the neutral potassium line at 7699A, resulted in a startling find: even prior to optical eclipse, the line strength was increasing, and increasing in a step-wise fashion. This trend continued into totality, thus far – see Figure 3. During last eclipse, Lambert & Sawyer (1986) had monitored this same line but their coverage was less frequent and did not reveal the wonderful details seen by Leadbeater.

Robin had been asking everyone in the campaign how to make sense of this stepwise behavior, and finally, during February 2010, it occurred to me that the slopes represented regions in the disk of enhanced density - "rings" separated by relative "gaps" - these I named A, B, C, D, E and F in their order of appearance during ingress. These density structures are being seen indirectly, by their effect on the *atmosphere of the disk -* wherein the neutral potassium line strength can be measured in the background F star light, assuming the disk itself to be dark and opaque, per the interferometric results. The word ring implies a circular symmetry, but we could easily be dealing with arcs or even spiral structure in the disk (see below). Details of this interpretation are available in an online paper by Leadbeater and Stencel (2010), and at Robin's homepage, http://www.threehillsobservatory.co.uk/ .

Figure 3: Variation of the excess equivalent width of neutral potassium in epsilon Aurigae during ingress.

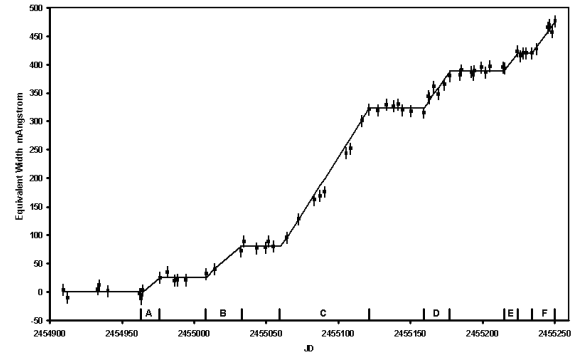

**The** appearance of the C ring coincides with the start of optical eclipse (RJD 55,060), and the gap between C and D coincides approximately with the ring gap discovered by Steno Ferluga (1990) and the brightening seen during ingress around RJD 55,150. Indeed, two of the steeper declines in light during ingress coincide with rings C and D, and the softening of light decline slope after each corresponds to neutral potassium line strength plateaus (gap CD and gap DE).

The C and D ring interior edges appear to coincide with local minima in the B and U filter light curves as well – coincidence? If anything, the C ring then CD gap correlate well with a redder then bluer U-B color, circa RJD 55150 (see photometry report by Jeff Hopkins, in the online newsletter series).

Disentangling the effect of rings and gaps from the light curve is an area of active investigation, given the overall complication of the "out of eclipse" (OOE) light variations arising from an uncertain but bright source in the system. It would be tempting to think that a series of semi-opaque clouds associated with disk structure could account for the OOE, but F star pulsation, or the production of photospheric hot spots could be responsible as well. The long term photometric record assembled by Lou Boyd and Jeff Hopkins does suggest that the amplitude of OOE variations are larger near times of eclipse and less at other times farthest from eclipse. Further study will be needed to determine the role of periastron passage, secondary minimum and/or circumbinary material in contributing to the OOE variations. Light curve differences from eclipse to eclipse support the idea of tidally raised and variable "spiral arm" structure in the disk, given similar effects seen in interacting galaxies and the face that periastron in this eccentric binary occurs not long before each eclipse. Estimating the survival of disk perturbations will require modeling. Note that the beta Pic debris disk was found to have ring like structure, perhaps as a result of a stellar encounter (Kalas et al. 2000). New Herschel Space Observatory far-IR observations confirm ring structure in Vega debris disk as well (Sibthorpe et al. 2010).

Another potential test of this ring structure hypothesis would be matching the polarization variations reported by Kemp et al. (1986). F star light "glinting" off of each ring-like cylinder could give rise to extra polarization through eclipse and out. Kemp's data started circa RJD 45200, which 9,890 days later, translates to RJD 55090, slightly after first contact. Kemp reported polarization in $Q = [I(0º) – I(90º)]$ and $U = [I(45º) – I(135º)]$, where I is the measured brightness at the given polarizer angle. Roughly speaking, as Q declined, U increased during several in-eclipse fluctuations.

Kemp reported polarization minima in Q near RJD 45300 and 45390, and polarization maxima near 45350 and 45440. These ingress minima translate to RJD 55190 [DE ring gap] and 55280 [local V band brightening 2010 March], current epoch. The maxima translate to 55240 [F ring] and 55330 [not yet observed]. Thus, we have preliminary support for these correlations, but will need to continue the monitoring to be certain. After mid-eclipse, Kemp shows a relative Q maximum near 45700, which is 55590, late Jan.2011. Also, computing spectro-polarization in H-alpha profiles from such geometry is feasible these days, and could provide additional confirmation.

One benefit of the neutral potassium work is the ability to make specific predictions for substructure transit times <u>during egress</u>, compared with previous eclipses and assuming persistence and symmetry of structures:
2010 Aug 4: Adopted mid-eclipse RJD 55413.
2010 Dec 19: Disc east inner rim 55550.
2011 Jan 14-30: F ring crossing 55576 – 55592.
2011 Feb 9-18: E ring crossing 55602 – 55611.
2011 Mar 19: 3rd contact, predicted 55640.
2011 Mar 28-Apr 15: D ring crossing 55649 – 55667.
2011 May 13: 4th contact, predicted 55695.
2011 May 23-Jul 24: C ring crossing 55705 – 55767.
2011 Aug 19-Sep 13: B ring crossing 55793 – 55818.
2011 Oct 15-28: A ring crossing 55850 – 55863.

During last eclipse, egress was very rapid, only 65 days, compared with 146 days recent ingress, per Jeff Hopkins. Should this difference recur, then either a symmetric C ring is not equally responsible for the first and fourth contacts, or the C ring is far from symmetric, relative to the disk center. The latter is feasible because as the disk rotates, the egress half of the disk is exposed to the heat of the F star, raising the disk temperature and sublimating materials. This effect has been seen in anomalous CO absorption appearing after mid-eclipse last cycle (Hinkle and Simon 1987). This changes the optical and infrared opacity, and will affect the light curves. Efforts are underway to repeat that series of near infrared spectra as well, and the photometry is equally well important.

## Development # 4: Theory

Two key questions still need to be addressed: the evolutionary status of the F star, and the dark disk's origin and fate. Together, the answers will form the basis for understanding the evolutionary status of this unique binary.

First, the F star: Given the SED and mass ratio, we face an Algol paradox – the less massive F star appears to be the more evolved one, with the luminosity of a supergiant, sustained by core fusion or in some very transient ballooned state, like a post-AGB star. However, the implication is that several solar masses of material were shed by the F star that perhaps was 6+ solar masses to begin with. But where is that missing mass? Neither IRAS or AKARI far infrared imaging detects any trail of dust left behind from a large mass loss episode.

When I suggest that a post-AGB star is present in the system, it doesn't necessarily imply that the star ever cooled down to an effective temperature like that of an M star (~3500K) - if it stayed hotter than G (~6000K), the UV output could have remained high enough to sublimate any dust, leaving only gas in the wake of the system - which would still be ionized, and an easy GALEX detection, in principle. One hypothesis is that little dust survived the UV radiation environment during mass loss in this system, except for that collected into the dusty disk itself. Here's a speculative, extreme view: because it is a binary, the now F star attempted to get rid of 3+ solar masses but much of it remained trapped as gas filling BOTH Roche lobes – and some collapsing into the disk around the B star, and some of it raining back down onto the now F star. Webbink (1985) discussed options like this in binary star evolution theory, but new calculations are needed. Because the orbital separation is so large, perhaps this special case occurs. As always, we need to find clever ways to test these ideas.

If the F star is in some hyper-extended post-AGB configuration, one could asteroseismic signatures to be present – large convective cells or global oscillations that contribute to the well-known but not understood out of eclipse light variations, on timescales of months. There is a hint of hot spots in the initial interferometric images, and tidal effects from the massive neighbor in an eccentric orbit might be important in sustaining F star envelope instabilities. This sort of study can be pursued even post-eclipse.

Second, the disk and its contents: As early as 1924, Ludendorff remarked that the eclipse could be caused by a "swarm of meteorites". Kopal (1954) introduced the notion of a large dark disk as the eclipse cause, and Huang (1964, 1974) elaborated this disk model to account for eclipse details. In a prescient paper, Takeuti (1986) describes the implications of the low mass model proposed by Saito, and describes the nature of a low mass disk heated by radiation from both stars in the binary. The balance of this eclipse will provide tests for Takeuti's ideas.

For perspective, given the binary star parameters, the outer portions of the dark disk has a Keplerian rotation period of nearly 3 years (Lissauer et al. 1996), implying that material in the disk faces the F star frequently (possibly resonantly) during the 27.1 year orbital period. Saito et al. (1987) asked whether a UV drive shock exists in the dawn quadrant of the disk? Their model addresses the gas component, most sensitive to ionizing radiation. One would need to compute a model for this, but intuitively the ionization would cause expansion of the gas, which would probably drag along some of the dust or further segregate dust in the plane from a gaseous envelope of the disk. The shell line spectrum that resembles the F star is undoubtedly from a source like this. If potassium and sodium are driven off dust particles by UV radiation - as is the case for Mercury and Io (Schaefer & Fegley, 2009) and perhaps in IRC 10 216 (Mauron & Huggins, 2010), they might still trace the underlying ring/spiral structure in the dusty disk. I'm hoping our next set of interferometric images can see into this portion of the disk/central clearing moreso, and map the second half of the disk in detail, thus perhaps help disentangle the role of some of these processes at work. Observations with MIRC at CHARA are scheduled to resume in late August 2010.

Does the disk in epsilon Aurigae resemble other so-called debris disks, like beta Pictoris and others? I think it makes a useful hypothesis for these reasons: First, there is a strong infrared excess relative to the F star photosphere - fitted by a 550K source (Hoard et al. 2010, Backman and Gillett, 1985). The IR luminosity of the disk implies a large surface area, that can be fit by an HHS disk, 0.95 AU thick and 7.8 AU long, as constrained by eclipse depth & duration. Thus, it is now established that the disk exists, as confirmed by interferometric imaging. Second, epsilon Aurigae presents a featureless IR spectrum: often interpreted as caused by a large dust grains in excess of 1 micron size. This has been recently confirmed via a series of spectra obtained with IRTF Spex (JHK, Stencel), BASS (10 micron, thanks to Michael Sitko), MIRAC4 (10 micron thanks to Joe Hora), plus my own Spitzer IRS spectra in 2005/06 used in the SED study. All of these show a smooth continuum and weak H recombination lines in emission - no sign of ices or molecules in either pre-eclipse spectra, or thus far among in-eclipse ones (thru mid-March 2010). I have asked for time to look for the CO lines at 2.3 microns expected to re-appear after mid-eclipse, and will re-propose to get MIRAC 10 micron spectra at the MMT/Whipple 8 meter to contrast the featureless spectra obtained there, in Jan 2010 in collaboration with Joseph Hora (CfA).

To a first approximation, YSO disks are still accreting ISM materials from their associated collapsing protostellar cores, and thus still show strong near-IR and mid-IR spectral features of ices, water and silicates. In contrast, debris disks have 'aged' enough for everything to condense into icy sub-planetessimal bodies (mm-km) and give rise to their featureless IR spectra. If so, then epsilon Aurigae more nearly resembles a debris disk case than a YSO type. Is there another class of disk that results from the mass transfer case? Perhaps. What are those distinguishing features of such a disk relative to YSO and debris cases? Obviously the UV irradiation has been 'toasting' the epsilon Aurigae disk for some time. In any event, all these questions point to a key parameter – the AGE of the system, relative to a mass transfer event. I suspect and hope to prove that this disk is dynamically young and there may be "a lot of junk" still moving about in the Roche lobes of both components. If we are fortunate to complete this eclipse with the continuation of as much fine data as has been seen to date, these questions will be better addressed.

To summarize, as we approach mid-eclipse, the following facts are emerging and challenge theory of evolution and binary interaction:
--the F star is overluminous for its inferred low mass;
--the existence of a disk appears certain, but details of its overall shape, composition, age and evolutionary status remain to be defined;
--the disk appears to possess an extended atmosphere, as detected in the neutral metals potassium and sodium plus other low temperature lines, and this gas may overfill the Roche lobe of the disk and its central star;
--the disk appears to have internal structure consistent with rings or tidally-induced spiral density waves, based on neutral potassium variations during ingress plus polarimetry from the last eclipse;
--ideally, mid-eclipse will enable an interferometric view of the center of the disk and its contents;
--ideally, egress will provide further spectroscopic evidence for the nature of dust in the disk, in terms of sublimation production of CO (and perhaps ices), caused by differential heating, revealing composition and dynamics. For all these reasons and more, we urge all observers to keep up the effort to provide as much detail in these coming phases as has been seen for ingress.

## Mid-eclipse and egress

Eclipse is forecast to continue throughout 2010 and well into spring 2011. Previous eclipses have shown persistence ***past optical egress*** in some indicators, like K I 7699A and the near IR CO bands. In addition to encouraging optical and spectroscopic monitoring to continue, here's hoping that telescope Time Allocation Committees will recognize the value of this singular observing opportunity, as epsilon Aurigae begins to conclude its first eclipse of the new millennium. Beyond the eclipse itself, additional perhaps robotic monitoring of the optical brightness of the system has merit to establish the persistence of previously noted post-eclipse variations. Insofar as possible, look for updates to appear at www.twitter.com/epsilon_Aurigae .

## Acknowledgements

It is a pleasure to acknowledge the many interested parties that have contributed either data or ideas to the ongoing study of epsilon Aurigae, much of which appears in some form in this paper. In addition to coauthors on other reports, a partial list includes: Jeff Hopkins, Don Hoard, Steve Howell, Brian Kloppenborg, Robin Leadbeater, Ed Guinan, Arne Henden, Joseph Hora, Dale Mais, Aaron Price, Mike Sitko, Don Terndrup, Rebecca Turner, Ryan Wyatt, and many others. This work is supported in part by NSF grant AST 10-16678 to the University of Denver, in addition to sustaining support from the bequest of William Herschel Womble in support of astronomy at the University of Denver, to whom this author is especially indebted.

## References

Backman, D. and Gillett, F., 1985 Ap.J. 299: L99. *Epsilon Aurigae during eclipse: IRAS observations.* http://adsabs.harvard.edu/abs/1985ApJ...299L..99B .

Carroll, S.,Guinan, E., McCook, G. and Donahue, R., *"Interpreting epsilon Aurigae"* (1991) Astrophys. Journal **367**, 278. http://adsabs.harvard.edu/abs/1991ApJ...367..278C .

Chadima, P., Harmanec, P., Yang, S., Bennett, P., Božić, H., Ruždjak, D., Sudar, D., Škoda, P., Šlechta, M., Wolf, M., Lehký, M. and Dubovský, P. 2010 IBVS 5937 "*A new ephemeris and an orbital solution of epsilon Aurigae*" - http://adsabs.harvard.edu/abs/2010IBVS.5937....1C .

Ferluga, S., 1990 Astron. Astrophys. 238: 270. *Epsilon Aurigae. I - Multi-ring structure of the eclipsing body.* http://adsabs.harvard.edu/abs/1990A%26A...238..270F .

Ferluga, S. and Mangiacapra, D., "*Epsilon Aurigae. II – The shell spectrum*" (1991) Astron. & Astrophys.

**243**: 230. http://adsabs.harvard.edu/abs/1991A%26A...243..230F .

Harrington, D. and Kuhn, J. "Spectropolarimetric Surveys: HAeBe, Be and Other Emission-Line Stars" (2009) Astrophys. Journal Supplement **180,** 138. http://adsabs.harvard.edu/abs/2009ApJ...695..238H .

Hoard, D., Howell, S. and Stencel, R., 2010 Ap.J. - in press (HHS). *Taming the invisible monster: system parameters for epsilon Aurigae from the far UV to the mid-IR.* http://arxiv.org/abs/1003.3694 .

Hopkins, J. L. Epsilon Aur Campaign Newsletters. 2010 http://www.hposoft.com/Campaign09.html .

Hopkins, J.L. And Stencel, R.E. 2009 "Epsilon Aurigae: A mysterious star system" (Book). ISBN 978-0-615-24022-0. http://www.hposoft.com/Campaign09.html .

Huang, S.S., 1965 "*An interpretation of epsilon Aurigae*." Astrophys. Journal **141**: 976. http://adsabs.harvard.edu/abs/1965ApJ...141..976H .

Huang, S.S., 1974 "*Interpretation of Epsilon Aurigae. III. Study of the light curve based on disk models.*" Astrophys. Journal **189**: 485. http://adsabs.harvard.edu/abs/1974ApJ...189..485H .

Kalas, P., Larwood, J., Smith, B. and Schultz, A. 2000 "*Rings in the Planetesimal Disk of beta Pictoris.*" Astrophys. J. 530: L133. http://arxiv.org/pdf/astro-ph/0001222v1

Kemp, J., Henson, G., Kraus, D., Beardsley, I., Carroll, L., Ake, T., Simon, T. and Collins, G., "*Epsilon Aurigae - Polarization, light curves, and geometry of the 1982-1984 eclipse*" (1986) Astrophys. Journal Letters **300**, L11. http://adsabs.harvard.edu/abs/1986ApJ...300L..11K .

Kloppenborg, B., Stencel, R., Monnier, J.D., Schaefer, G., Zhao, M., Baron, F., McAlister, H. A., ten Brummelaar, T. A., Che, X., Farrington, C.D., Pedretti, E., Sallave-Goldfinger, P.J., Sturmann, J., Sturmann, L., Thureau, N., Turner, N., and Carroll, S., 2010 Nature Letters 464, 870. *Infrared images of the transiting disk in the epsilon Aurigae system.* *http://adsabs.harvard.edu/abs/2010Natur.464..870K*

Kopal, Z. 1954 "*The nature of the eclipses of epsilon Aurigae*" The Observatory **74**: 14. http://adsabs.harvard.edu/abs/1954Obs....74...14K .

Lambert, D. and Sawyer, S. "*Epsilon Aurigae in eclipse. II - Optical absorption lines from the secondary*" 1986, Publications Astronomical Society of the Pacific **98**: 389. http://adsabs.harvard.edu/abs/1986PASP...98..389L .

Leadbeater, R. and Stencel, R. 2010, "*Structure in the disc of epsilon Aurigae: evidence from spectroscopic monitoring of the neutral potassium line during eclipse ingress*" - http://arxiv.org/abs/1003.3617 .

Lissauer, J., Wolk, S., Griffith, C. and Backman, D., 1996 Astrophys. J., 465, 371-384. *The epsilon Aurigae secondary: a hydrostatically supported disk.* *http://adsabs.harvard.edu/abs/1996ApJ...465..371L*

Mauron, N. and Huggins, P. 2010 Astron. Astrophys. 513: 31. "Gas phase atomic metals in the circumstellar envelope of IRC+10216" http://arxiv.org/pdf/1002.3554v1

Nha, I., Lee, Y., Jeong, J. and Kim, H., "*Extreme Long-period Eclipsing Binary epsilon Aurigae*" (1993), in Proceedings New frontiers in binary star research, eds. Kam-Ching Leung and Il Seong Nha, ASP Conf. Vol. **38**, p.291. http://adsabs.harvard.edu/abs/1993ASPC...38..291N .

Nordgren, T., Sudol, J. and Mozurkewich, D. "*Comparison of Stellar Angular Diameters from the NPOI, the Mark III Optical Interferometer, and the Infrared Flux Method*" - 2001, Astronomical Journal, **122**, 2707. http://adsabs.harvard.edu/abs/2001AJ....122.2707N .

Schaefer, L. And Fegley, B. 2009 Astrophys. J. 703, L113. "*Chemistry of silicate atmospheres of evaporating super-Earths*" http://arxiv.org/pdf/0906.1204v1 .

Sibthorpe, B. and 39 coauthors, 2010 - "The Vega Debris Disc: A View from Herschel" http://adsabs.harvard.edu/abs/2010arXiv1005.3543S .

Stefanik, R., Lovegrove, J., Pera, V., Latham, D., Torres, G. and Zajac, J. "*Epsilon Aurigae: An Improved Spectroscopic Orbital Solution*" (2010) Astron. J. 139, 1254. http://arxiv.org/abs/1001.5011 .

Stencel, R. 1985, The 1982-84 Eclipse of epsilon Aurigae - NASA Conf. Publication No. 2384 – weblink:
[ http://www.du.edu/~rstencel/NASAcp2384.pdf ].

Stencel, R., Creech-Eakman, M., Hart, A., Hopkins, J., Kloppenborg, B., Mais, D., "*Interferometric Studies of the Extreme Binary epsilon Aurigae: Pre-Eclipse Observations*" (2008) Astrophys. Journal **689**, L137.
http://adsabs.harvard.edu/abs/2008ApJ...689L.137S .

Stencel, R. and Hopkins, J. (2009) "*Epsilon Aurigae, 2009: The eclipse begins – observing campaign status*" in Proceedings of the Society for Astronomical Sciences 28th Annual Symposium on Telescope Science, May, 2009.
http://adsabs.harvard.edu/abs/2009SASS...28..149S .

Struve, O. and Elvey, C., (1930) "*Preliminary Results of Spectroscopic Observations of 7 epsilon Aurigae*" Astrophys. Journal 71: 136.
http://adsabs.harvard.edu/abs/1930ApJ....71..136S .

Takeuti, M. 1986 "*An accretion disc surrounding a component of Epsilon Aurigae*" Astrophysics and Space Science 121: 127.

Webbink, R. 1985 "*Epsilon Aurigae in an Evolutionary Context*" in The 1982-84 Eclipse of epsilon Aurigae (ed. R. Stencel) NASA Conf. Publication No. 2384 – weblink:
[ http://www.du.edu/~rstencel/NASAcp2384.pdf ].

[end]